# Evaluation of hydrogen diffusion and trapping in ferritic steels containing (Ti,Cr)C particles using electrochemical permeation and thermal desorption spectroscopy

Nicholas Winzer, thyssenkrupp Steel Europe AG, Kaiser-Wilhelm-Straße 100, 47166 Duisburg, Germany, E-Mail: nicholas.winzer@thyssenkrupp-steel.com

**Abstract**

Hydrogen diffusion and trapping in ferritic steels containing (Ti,Cr)C particles was investigated using electrochemical permeation (EP) and thermal desorption spectroscopy (TDS). The trapping parameters for the test materials were evaluated by fitting the measurements with a finite element model based on the McNabb-Foster equations using least-squares optimisation. The measurements showed that hydrogen diffusion in ferrite is slowed significantly by the presence of fine (<5 nm) (Ti,Cr)C particles; coarser particles had little or no effect. The TDS measurements were consistent with hydrogen traps with a high energy barrier. The uniqueness of the hydrogen trapping parameters obtained using the fitting procedure was evaluated. It was found that the system was overdetermined; the measurements could be fitted with multiple combinations of trapping parameters. Consequently, it was not possible to determine the individual trapping parameters using this procedure. Trapping parameters were also evaluated from TDS measurements by applying Kissinger's equation. Using this procedure a trap binding energy of 0.24 eV was calculated for all materials, albeit with a high degree of uncertainty.

## 1    Introduction

Hydrogen diffusion and trapping are key factors affecting the susceptibility of metals to hydrogen embrittlement (HE). The experimental evaluation of hydrogen diffusion and trapping is, therefore, essential to understanding the susceptibility of metals and metallic components to HE. The most commonly-used experimental methods for evaluating hydrogen diffusion and trapping in metals are electrochemical permeation (EP) and thermal desorption spectroscopy (TDS). However, despite their widespread use there is still disagreement regarding the interpretation of EP and TDS measurements. This has led to uncertainty regarding the validity of experimental results for some materials.

First-order evaluation of EP and TDS measurements can be achieved using analytical equations. In the case of EP measurements, this usually involves calculating the effective diffusion constant, $D_{eff}$, based on the lag time (time until the measured hydrogen flux reaches some threshold) or by fitting the permeation curve with the analytical solution to Fick's second law [1-11]. This approach provides little information regarding the underly traps. It is also contingent on the measured curve closely resembling the analytical solution to Fick's second law, which for complex materials is often not the case. The most common approach to evaluating TDS measurements is based on the Kissinger equation [2,4,8,10,12-38]:

$$\frac{\partial \ln (\varphi/T_M^2)}{\partial (1/T_M)} = -E_D/R \qquad \text{Equation 1}$$

where $\varphi$ (K·s$^{-1}$) is the heating rate, $E_D$ (eV) is the activation energy for hydrogen desorption, $R$ (eV·K$^{-1}$·atom$^{-1}$) is the ideal gas constant and $T_M$ (K) is the temperature corresponding to the maximum hydrogen flux. Thus $E_D$ can be determined by plotting $\ln (\varphi/T_M^2)$ against $1/T_M$ for various values of $\varphi$. However, this approach assumes that desorption is not affected by diffusion in normal lattice sites, which may lead to erroneous results for thick samples or materials with low lattice diffusivity.

A more rigorous approach to evaluating EP and TDS measurements involves fitting the experimental curves with a finite element (FE) model based on equations for hydrogen diffusion in the presence of traps using optimisation algorithms [6,39-50]. Variations in this approach are mostly due to differences in the underlying diffusion equations. A generalised set of equations for hydrogen diffusion in the presence of saturable traps was given by McNabb and Foster [51]:

$$\frac{\partial C_L}{\partial t} = D_L \nabla^2 C_L - \sum_{i=1}^{n} N_i \frac{\partial \theta_i}{\partial t} \qquad \text{Equation 2}$$

Where:

$$\frac{\partial \theta_i}{\partial t} = \kappa_i C_L (1 - \theta_i) - \lambda_i \theta_i \qquad \text{Equation 3}$$

Where $C_L$ (atoms·m$^{-3}$) is the concentration of hydrogen in normal interstitial lattice sites, $D_L$ (m$^2$·s$^{-1}$) is the diffusivity of hydrogen in normal interstitial lattice sites (i.e., in the absence of traps) and $N_i$ (sites·m$^{-3}$) and $\theta_i$ are the density and occupancy of trap $i$ respectively. $\theta_i$ is given by $C_i/N_i$, where $C_i$ (atoms·m$^{-3}$) is the concentration of hydrogen in trap $i$. $\kappa_i$ (m$^3$·s$^{-1}$) and $\lambda_i$ (s$^{-1}$) are the rate constants associated with trapping and de-trapping for trap $i$ respectively and are given by [52]:

$$\kappa_i = \frac{\nu}{N_L} \exp\left(\frac{-E_{\kappa,i}}{RT}\right) \qquad \text{Equation 4}$$

$$\lambda_i = \nu \exp\left(\frac{-E_{\lambda,i}}{RT}\right) \qquad \text{Equation 5}$$

where $\nu$ (s$^{-1}$) is the is the oscillation frequency of hydrogen, $N_L$ (sites·m$^{-3}$) is the density of normal interstitial lattice sites and $E_{\kappa,i}$ and $E_{\lambda,i}$ (both eV) are the activation energies for trapping and de-trapping respectively, as shown schematically in Figure 1. Also shown in Figure 1 is the activation energy for diffusion in normal lattice sites $E_L$ (eV) and the trap binding energy $\Delta E_{T,i}$ (eV), which is given by $E_{\kappa,i} - E_{\lambda,i}$.

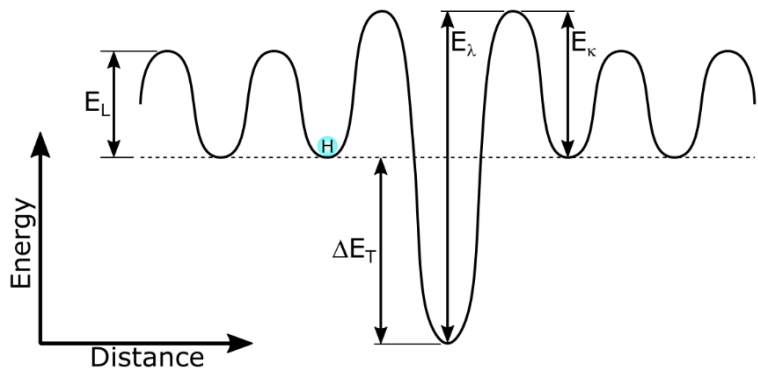

Figure 1 – Schematic illustration of the binding energies for hydrogen diffusion and trapping.

FE models based on the McNabb-Foster equations have been widely used in the literature to evaluate EP [39-42] and TDS [41,43-49] measurements. Winzer et al. [42] used a FE model based on the McNabb-Foster equations to fit EP measurements for advanced high strength steels; however, it was shown that in most cases the combinations of parameters obtained using this approach are not unique.

Oriani [53] proposed a simplification to the McNabb-Foster equations based on the assumption of local equilibrium between hydrogen in traps and in the surrounding lattice sites. In this case the trap energetics are described solely by $\Delta E_{T,i}$. Consequently, the energy barriers, $E_{\kappa,i}$ and $E_{\lambda,i}$, cannot be evaluated using this approach. FE models based on the Oriani approximation have also been used in the literature to evaluate EP [6] and TDS [48,50,54,55] measurements.

The use of titanium carbide (TiC) particles to mitigate HE of steels has long been a focus of investigation [27,56-63]. It is generally accepted that TiC particles may act as hydrogen traps, with the trapping characteristics of TiC particles dependent on the coherency of the interface with surrounding lattice [64-67] and the charging temperature [68-70]. There is little consensus regarding the exact trapping sites at TiC particles. It has been proposed that hydrogen trapping occurs in the surrounding strain field [71-74], the particle interface [54,55,64,65,67,71-78], misfit dislocations in the surrounding matrix [64,67,68,71,72,79] or at vacancies at the interface [54,55,64,65,68,71,72,74,78,79,80] or in the particle core [54,55,64,67,70-74,78,79-84]. Correspondingly a broad range of binding energies have been reported in the literature, as shown in the recent review by Vandewalle *et al.* [85]. Recently there has also been increased interest in mixed carbides, such as (Ti,Mo)C, which purportedly have an increased ability to trap hydrogen at room temperature [72,76-78,86,87].

The goal of the present study was to evaluate hydrogen trapping in ferritic steels by (Ti,Cr)C in steels using EP and TDS measurements. The experimental measurements were evaluated using a FE model based on the McNabb-Foster equations in order to capture as much detail as possible regarding the trap energetics. The uniqueness of the results was evaluated by investigating the correlations between the resulting trapping parameters.

## 2    Materials and Methods

### 2.1    Materials

The material used in this study was a hot rolled ferritic steel with the nominal composition: <0.1 wt.-% C; <2.0% wt.-% Mn; <0.2 wt.-% Ti; and <0.2 wt.-% Cr. The steel was cast in the form of 100 kg ingots in a laboratory-scale vacuum induction melting furnace. The cast ingots were rolled to a thickness of 50 mm and machined into blocks, which were then wrapped in foil and solution annealed at 1250 °C for 1 hr. The blocks were subsequently hot rolled to a thickness of 2.5 mm, water cooled to the coiling temperature and transferred to a furnace, where they were held at the coiling temperature for 1 hr and subsequently cooled at 0.5 °C/hr to room temperature. The coiling temperatures, which are denoted T1 – T5 in order of increasing temperature, were adjusted to achieve microstructures with and without TiC particles and with particles of varying sizes. At T1 no particles were precipitated. At temperatures T2 – T5 the particles were precipitated and underwent varying degrees of coarsening. The resulting materials are referred to herein according to the coiling temperature.

### 2.2    Microstructural analysis

Microstructural analysis was performed using optical microscopy (OM) and transmission electron microscopy (TEM). OM samples were polished using OPS colloidal silica and etched using nital (4% $HNO_3$ in ethanol). Samples for TEM were prepared using the carbon microprint method.

## 2.3 Electrochemical permeation tests

EP tests were performed according to ISO 17081 [88] using a Davanathan-Stachurski [89] permeation cell made from glass. The samples were ground to a thickness of 2 mm to remove surface defects. Immediately prior to testing the samples were again ground with 320 grade emery paper and degreased in ethanol. On both sides of the sample the diameter of the exposed surface was 3.5 cm. On the cathodic side the sample was charged in 0.05 mol/L $H_2SO_4$ at a constant current density of 120 $A/m^2$ using a platinum-coated counter electrode and a Yokogawa GS200 current source. On the anodic side the sample was polarised in 0.1 mol/L NaOH at +300 mV relative to a saturated calomel electrode using an IPS 10V-2A-E potentiostat. All tests were performed at 30 °C. For each sample two consecutive tests were performed. After the first test the sample was removed from the cell and left at room temperature for at least 24 hours before performing the second test.

## 2.4 Thermal desorption tests

TDS measurements were performed using a Bruker IR07 infrared furnace with a InProcess Instruments ESD100 mass spectrometer. Samples of dimensions 20 x 100 mm were ground to the same thickness and surface finish as the permeation samples and degreased in ethanol prior to hydrogen charging. Hydrogen charging was performed in 0.5 mol/L $H_2SO_4$ + 1.3×$10^{-5}$ mol/L $NH_4SCN$ at a constant current density of 10 $A/m^2$ using a platinum-coated counter electrode and a Yokogawa GS200 current source. The charging time for the different materials varied between 3 and 18 hours according to the time required to reach steady stated conditions in the EP tests. Immediately after charging the samples were cut in half and immersed in liquid nitrogen until the TDS measurements could be performed. Immediately prior to performing the TDS measurements the samples were cleaned and warmed to room temperature in acetone. The samples were heated in a quartz tube in a stream of $N_2$ gas at heating rates between 0.12 and 0.4 K/s whilst the desorbed hydrogen was measured using the mass spectrometer. The temperature of the sample was measured using a thermocouple placed on the centre of its broad surface. All measurements were performed up to a temperature of at least 1000 K; however, for all materials no significant peaks were detected at temperatures greater than 600 K. It should be noted that the quartz tube in which the sample was heated was open to the laboratory environment. Thus, measurements at higher temperatures may be affected by background hydrogen, e.g. in water contained in the atmosphere or adsorbed onto the walls of the quartz tube. For these reasons only desorption peaks up to temperatures of 600 K have been considered. The desorption flux at the onset of heating was subtracted from the measured flux at all temperatures, such that the initial desorption flux was zero. A linear correction was also applied up to the end of the initial peak to account for drift in the measured signal due to background hydrogen.

## 2.5 Details of model

Simulation of EP and TDS measurements were performed using the Python application FESTIM [43-45]. This solves the McNabb-Foster [51] equations for hydrogen diffusion in the presence of traps using the finite element method. For most simulations the samples were discretised using 100 unidimensional linear elements of equal length. For simulations of EP and TDS measurements step sizes of 100 and 5 s were used respectively in most cases.

For the EP measurements the initial value of $C_L$ was assumed to be zero. The lattice concentration at the charging surface, $C_{L,x=0}$ (atoms·$m^{-3}$), was assumed to be constant and equal to:

$$C_{L,x=0} = \frac{i_{max}X}{E_C D_L} \qquad \text{Equation 6}$$

where $i_{max}$ is the steady-state current density (A·m$^{-2}$), $X$ is the sample thickness (m) and $E_C$ is the elementary charge (C). The lattice concentration at the anodic surface, $C_{L,x=L}$ (atoms·m$^{-3}$), was assumed to be zero at all times. The current density on the anodic side of the sample was calculated using:

$$i_{x=L} = E_C F \qquad \text{Equation 7}$$

where $F$ (atoms·m$^{-2}$·s$^{-1}$) is the flux through the anodic surface.

For the TDS measurements the initial total hydrogen concentration was calculated by integrating the measured TDS spectra. The temperature was taken from the measured value using a modification to the FESTIM code [90]. It was assumed that all hydrogen was contained in traps; any hydrogen in lattice sites would have escaped rapidly prior to onset of the temperature ramp. $C_L$ was assumed to be zero at both surfaces at all times. The total desorption flux was taken as the sum of the fluxes through both surfaces.

$D_L$ was taken to be equal to $7.32 \times 10^{-8} \exp(-E_L/RT)$ where the $E_L$ is the activation energy for hydrogen jumping between normal interstitial lattice sites and was taken to be 0.059 eV [91]. $\nu$ was taken to be equal to $10^{13}$ s$^{-1}$ as proposed by Legrand *et al.* [92] and in line with ab-initio calculations for hydrogen in BCC Fe by others [93,94]. $N_L$ was taken to be 5.2 x 10$^{29}$ sites·m$^{-3}$, which corresponds to the density of tetrahedral sites in BCC Fe [95].

Fitting was performed using the Python library lmfit [96], which is based on the Levenberg-Marquardt algorithm. In all examples given in this report a single type of trap was assumed. Unless otherwise stated the trapping parameters $N$, $E_\kappa$ and $E_\lambda$ were optimised to fit the model with the experimental measurement.

## 3 Results

### 3.1 Microstructural analysis

Figure 2 shows optical micrographs of the materials T1 – T5. The microstructure of the material T1 comprised fine bainitic ferrite grains, whereas those of the materials T2 – T5 comprised coarse polygonal ferrite grains with some traces of pearlite. Coarse titanium carbonitride particles were present in all materials.

TEM images of the microstructures of the test materials are shown in Figure 3. The microstructure of material T1 was largely featureless. This suggested that carbides were not precipitated at temperature T1. In contrast, T2 – T5 contained a large number of fine particles. The size of the particles increased with increasing coiling temperature. For T2 and T3 the particles were relatively uniform and less than 5 nm in diameter. For T4 and T5 the sizes of the particles varied, consistent with Ostwald ripening, with the average diameter greater than 10 nm. EDX analysis of the particles showed that they had the nominal composition (Ti,Cr)C.

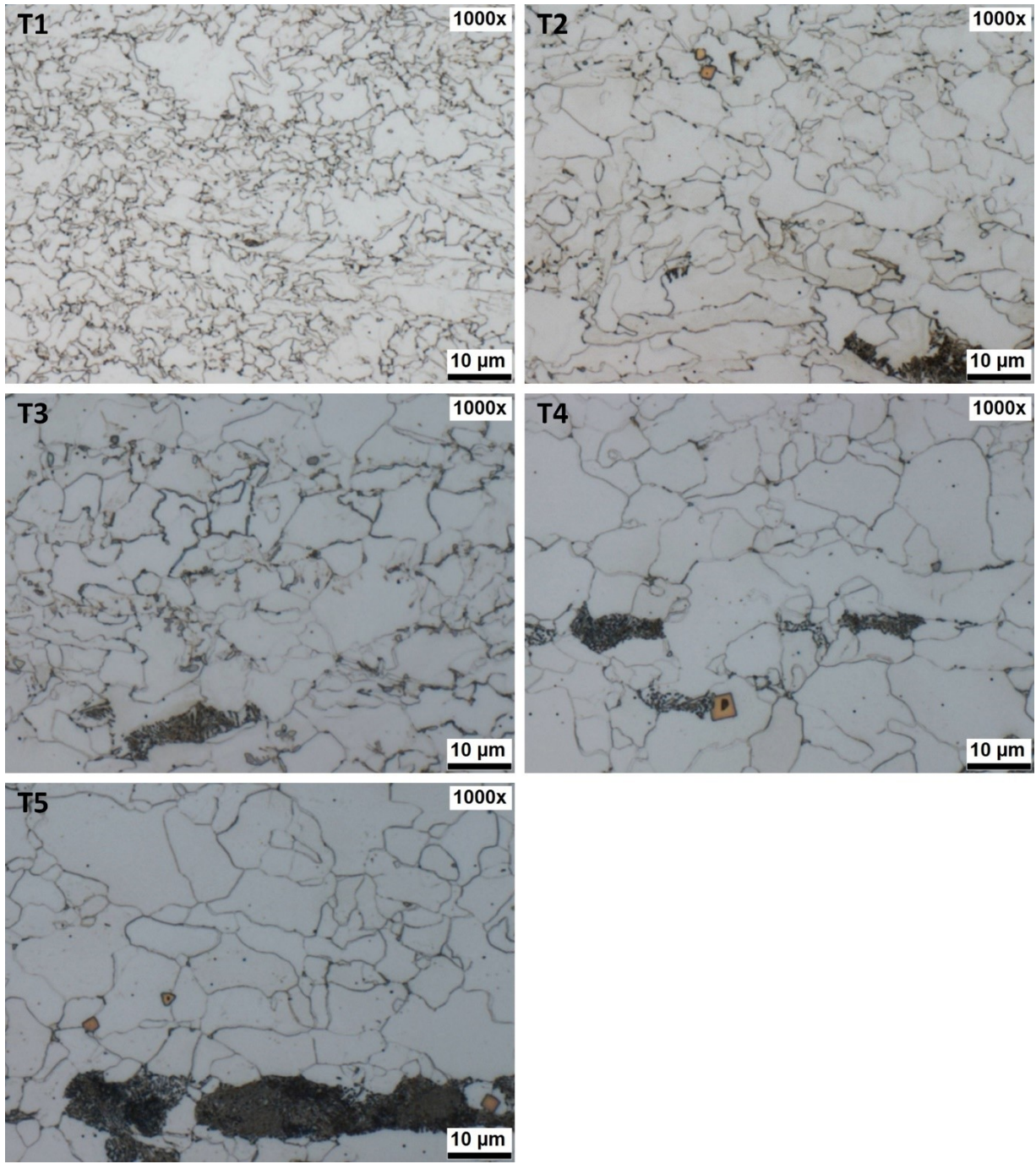

Figure 2 – Optical micrographs of the materials produced by simulated coiling at temperatures T1 – T5.

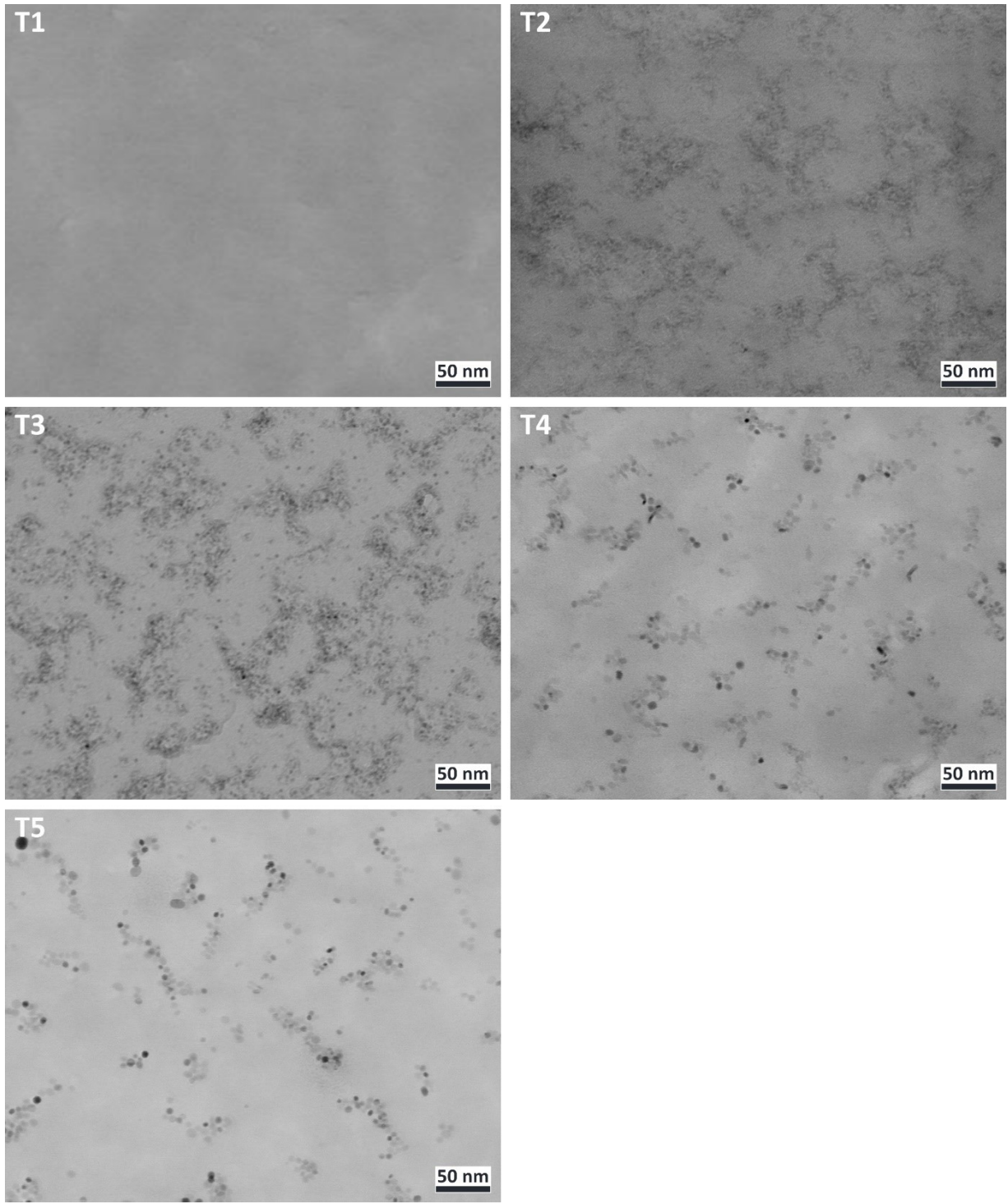

Figure 3 – TEM images of the materials produced by simulated coiling at temperatures T2 – T5. The dark spots correspond to (Ti,Cr)C particles.

*3.2 EP measurements*

Figure 4 shows examples of EP measurements of the test materials normalised with respect to the steady-state current, $I_{max}$. For T5 the time to reach steady state was less than that for the other materials. This indicates that the hydrogen diffusivity of T5 was higher than that of the other materials. A reduction in coiling temperature corresponded with a reduction in hydrogen diffusivity, characterised by an increase in the time to reach steady state; the diffusivities of T2 and T3 were lower than those of T4 and T5. Further reduction of the coiling temperature to T1 corresponded with a decrease in time to reach steady state, or an increase in diffusivity, relative to T2 and T3. The results of the EP measurements correlate with the presence and size of (Ti,Cr)C particles as shown in Figure 3. For T2 – T5 the hydrogen diffusivity increased with decreasing particle size. For T1, which did not contain particles, the diffusivity was similar to that of material T4. For T1, T4 and T5 there was no significant difference between the first and second permeation curves. In contrast, for T2 and T3 the diffusivity was higher for the second test than for the first test. This is consistent with the presence of irreversible trapping sites in these materials.

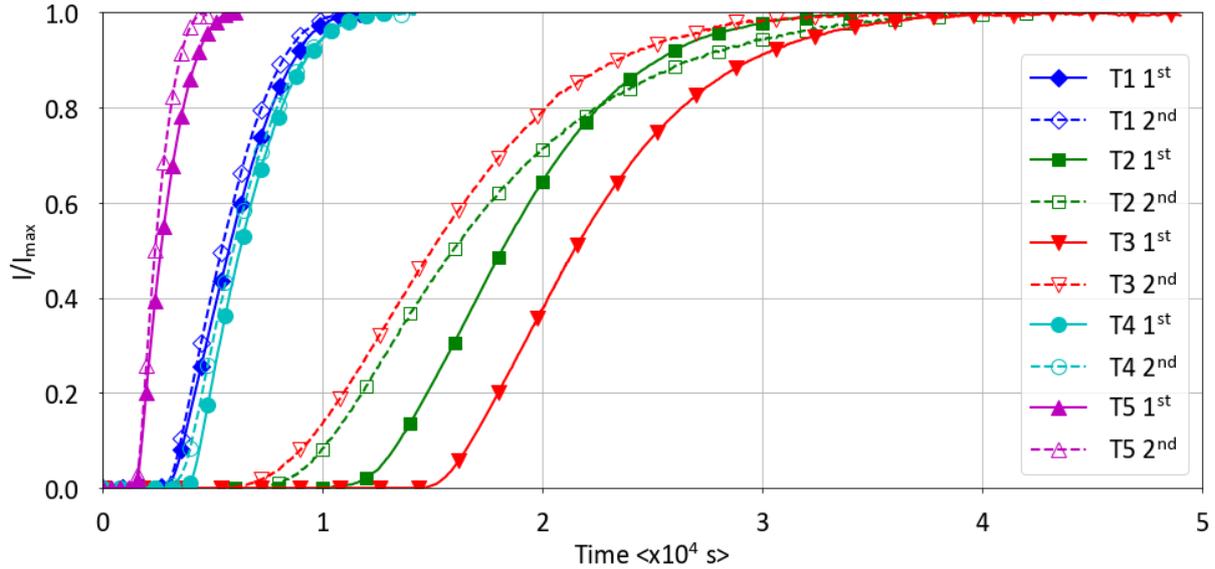

Figure 4 – Measured first and second EP curves for materials T1 – T5 normalised with respect to the steady-state current, $I_{max}$.

Figure 5 shows an example of an EP measurement (first test) for T3 as well as the analytical solution to Fick's second law, which was calculated using the equation [97-99]:

$$I = I_{max}\left[1 + 2\sum_{n=1}^{\infty}(-1)^n \exp\left(\frac{-D_{eff}n^2\pi^2 t}{X^2}\right)\right] \quad \text{Equation 8}$$

where $D_{eff}$ (m$^2$·s$^{-1}$) was calculated using the equation:

$$D_{eff} = \frac{X^2}{6t_{lag}} \quad \text{Equation 9}$$

where $t_{lag}$ (s) is determined using the procedure proposed by Winzer et al. [42], in which the cumulative current is plotted over time and $t_{lag}$ is taken as the intersect of the asymptote of the curve with the horizontal axis. For all materials there was a significant deviation between the measured EP curve and the analytical solution to Fick's second law for the first and second measurement. This indicates that hydrogen diffusion through the sample cannot accurately be described using $D_{eff}$.

Also shown in Figure 5 are two theoretical EP curves that were determined by fitting the experimental measurement with the FE model assuming a single type of trap. For all materials a single type of trap was sufficient to achieve a good fit with the first and second permeation measurements. For the theoretical curves shown in Figure 5 the optimum values of $E_\kappa$ and $E_\lambda$ were determined assuming $N$ = 1.4×10$^{24}$ m$^{-3}$ and 1.9×10$^{24}$ m$^{-3}$. The corresponding values for $E_\kappa$ and $E_\lambda$ and, as well as $\Delta E_T$, are given in Table 1. The two solutions have different values for $E_\kappa$ and $E_\lambda$, whereas $\Delta E_T$ is relatively consistent. This indicates that there exists no unique combination of $N$, $E_\kappa$ and $E_\lambda$; the measured curves may be fitted using multiple combinations of parameters.

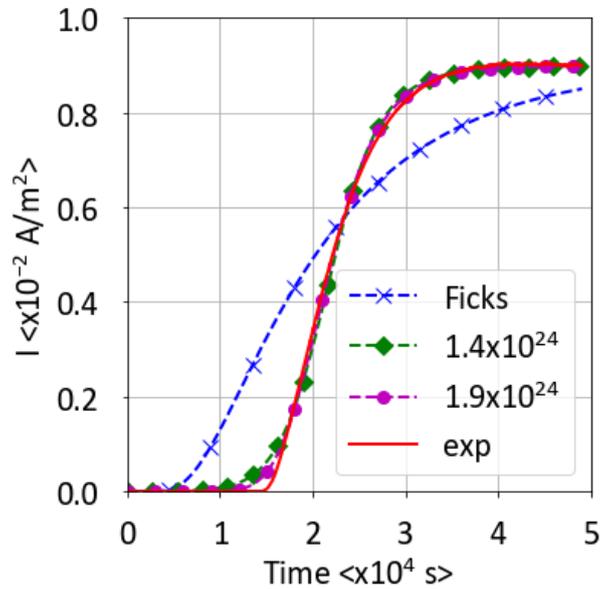

Figure 5 – Measured EP measurement (first test) for T3 (solid red line) and theoretical curves obtained using different trap parameters. The parameters for the theoretical curves were obtained by assuming $N$ = 1.4×10$^{24}$ m$^{-3}$ (◆) or N = 1.9×10$^{24}$ m$^{-3}$ (●) and finding the optimal combinations of $E_\kappa$ and $E_\lambda$ (given in Table 1) using the fitting procedure. Also shown is the analytical solution to Fick's second law assuming a constant $D_{eff}$ (✗).

Table 1 – Values of $E_\kappa$ and $E_\lambda$, as well as $\Delta E_T$, obtained by fitting the numerical model to the measured EP curves shown in Figure 5 assuming fixed values for $N$.

| $N$ (x10$^{24}$ m$^{-3}$) | $E_\kappa$ (eV) | $E_\lambda$ (eV) | $\Delta E_T$ (eV) |
|---|---|---|---|
| 1.4 | 0.4609 | 1.017 | 0.5561 |
| 1.9 | 0.1244 | 0.6351 | 0.5107 |

*3.3 TDS measurements*

Figure 6 shows TDS measurements for all materials tested at a heating rate of 0.28 K·s$^{-1}$. All measurements were characterised by a single peak with the maximum desorption flux occurring at a temperature in the range 350 – 400 K. The maximum desorption flux and the corresponding temperature correlated with the presence and size of (Ti,Cr)C particles (see Figure 3) and with the EP measurements (see Figure 4). For T2 – T5 the decrease in particle size correlated with an increase in the maximum desorption flux and the corresponding temperature. The TDS measurements of T1, which did not contain (Ti,Cr)C particles, were similar to those of T4.

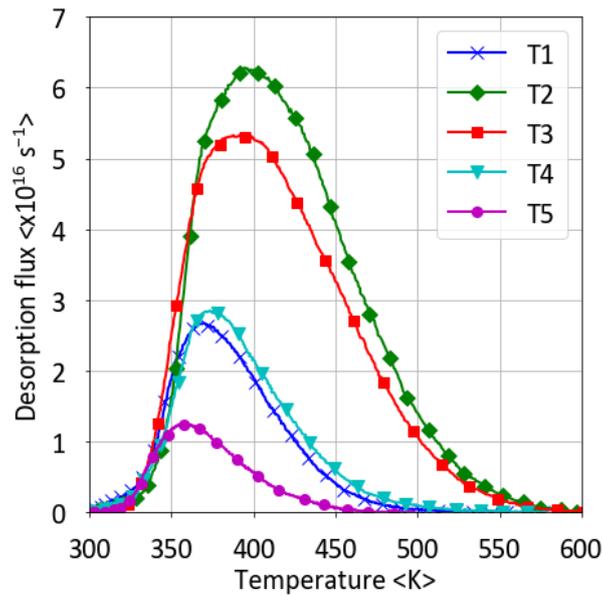

Figure 6 – TDS measurements for the materials T1 – T5 tested at a heating rate of 0.28 K·s$^{-1}$.

Figure 7A shows an example of a TDS measurement for T4 at a heating rate of 0.24 K·s$^{-1}$. Also shown in Figure 7A are two theoretical curves that were calculated using the FE model assuming a single type of trap. The parameters for the theoretical curves were determined by assuming $N$ = 10$^{25}$ m$^{-3}$ and 10$^{27}$ m$^{-3}$ and finding values of $E_\kappa$ and $E_\lambda$ to achieve the best fit with the measured curve using the fitting procedure. This process was repeated for values of $N$ from 5×10$^{24}$ m$^{-3}$ – 5×10$^{27}$ m$^{-3}$; for values of $N$ outside this range no solution could be found that achieved a good fit with the measurement. The corresponding values of $E_\kappa$ and $E_\lambda$, as well as $\Delta E_T$, are shown in Figure 7B with respect to $\ln(N)$. All combinations of parameters shown in shown in Figure 7B resulted in a good fit with the experimental measurement. For all combinations of parameters $E_\lambda$ is relatively consistent whereas $E_\kappa$ increases linearly with increasing $\ln(N)$. Consequently there is a negative linear correlation between $\Delta E_T$ and $\ln(N)$, with the gradient of $\Delta E_T$ vs $\ln(N)$ equal to -0.033 eV. The same correlation also occurred for the other materials, with the gradient of $\Delta E_T$ vs $\ln(N)$ in the range -0.03 – 0.045 eV. This indicates that within this range of $N$ there is no unique combination of $N$, $E_\kappa$ and $E_\lambda$ that will achieve a good fit with the measured curve.

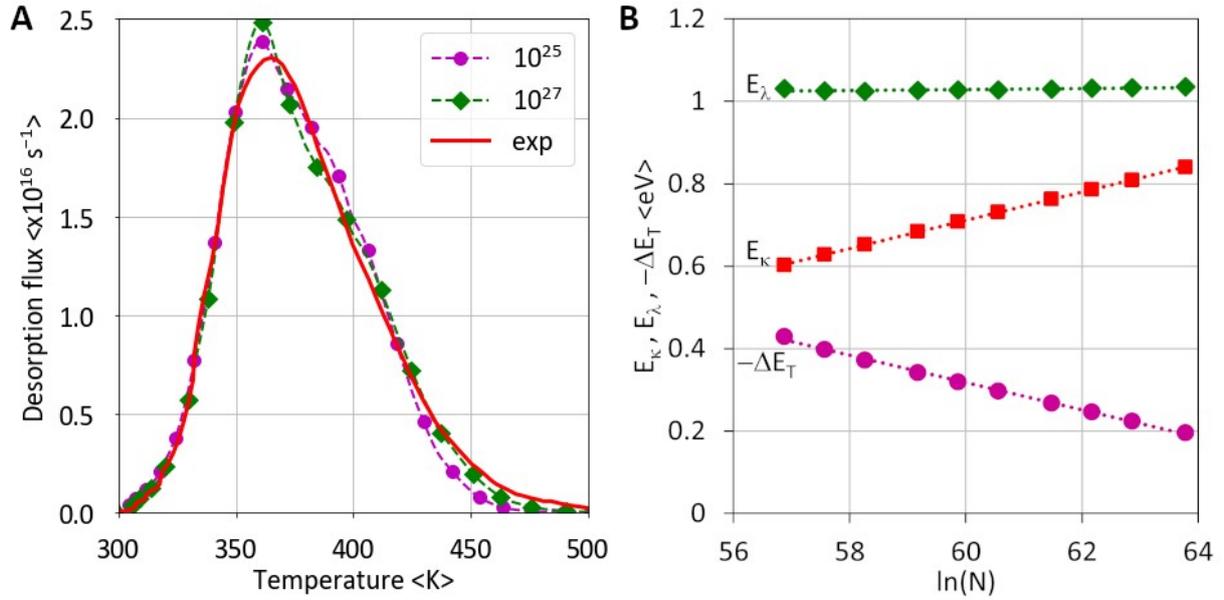

Figure 7 – (A) TDS measurement for T4 at a heating rate of 0.24 K·s$^{-1}$ (solid red line) and theoretical curves obtained by assuming $N = 10^{25}$ m$^{-3}$ (●) and $N = 10^{27}$ m$^{-3}$ (◆) and finding the optimal combinations of $E_\kappa$ and $E_\lambda$ using the fitting procedure. (B) Optimised values of $E_\kappa$ (■) and $E_\lambda$ (◆), as well as $\Delta E_T$ (●), obtained by fitting the numerical model to the experimental measurement shown in Figure 7A assuming various values of $N$ between $5\times10^{24}$ m$^{-3}$ and $5\times10^{27}$ m$^{-3}$.

### 3.4 Correlation between EP and TDS measurements

The fitting procedure was applied to all EP (first tests) and TDS measurements for all materials. For each measurement the optimum values for $N$, $E_\kappa$ and $E_\lambda$ were evaluated assuming a single type of trap. For the EP measurements, as shown in the example in Figure 5 and Table 1, the values for $E_\kappa$ and $E_\lambda$ varied; $E_\kappa$ was in the range 0.10 – 0.46 eV (Ave = 0.21, σ = 0.12) whereas $E_\lambda$ was in the range 0.60 – 0.96 eV (Ave = 0.70, σ = 0.13). $\Delta E_T$ was relatively constant and in the range 0.45 – 0.52 eV (Ave = 0.49, σ = 0.02). In contrast, for the TDS measurements the values for $E_\kappa$ and $E_\lambda$ were higher and less scattered than those for the EP measurements; $E_\kappa$ was in the range 0.38 – 0.70 eV (Ave = 0.59, σ = 0.05) whereas $E_\lambda$ was in the range 0.87 – 1.11 eV (Ave = 1.04, σ = 0.05). There was no correlation between the heating rate and the trapping parameters. Figure 8 shows the $\Delta E_T$ with respect to $\ln(N)$ for all materials. For the EP measurements the values for $\ln(N)$ were higher, and those for $\Delta E_T$ lower, than those for the TDS measurements. Evaluation of the TDS measurements for T1 and T5 resulted in a broad range of values. In contrast the values for T2, T3 and T4 were relatively consistent. For materials T1, T4 and T5 the values for the EP and TDS measurements were colinear. The gradients of the trend lines through all points were in the range -0.026 – -0.032 eV. It should be noted that the deviation between the theoretical and measured TDS curves varied; in some cases the deviation was considerably greater than that of the example shown in Figure 7A. There was no consistent influence on the deviation of the material, heating rate or trapping parameters. Thus, the deviation is likely a result of experimental variation and does not affect the relation between $\Delta E_T$ and $\ln(N)$.

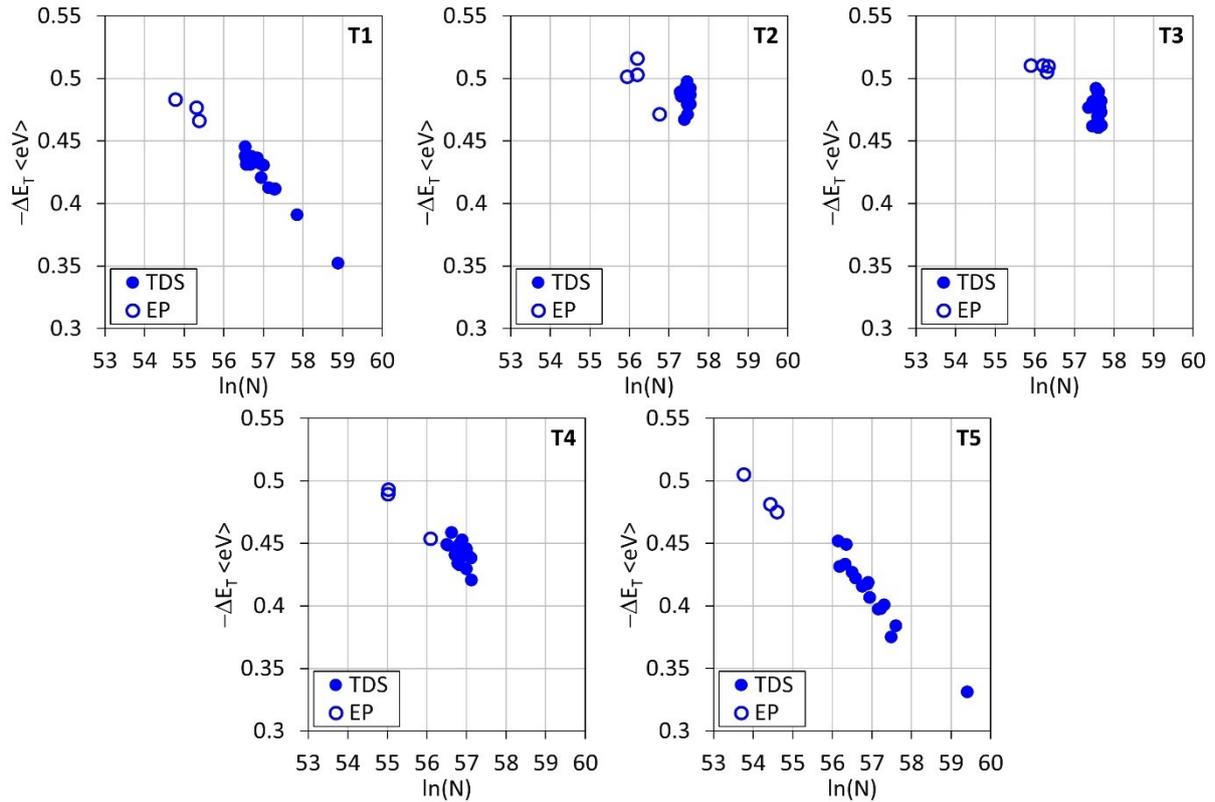

Figure 8 – Combinations of $N$ and $\Delta E_T$ obtained by fitting the numerical model to all EP (first tests) and TDS measurements for the materials T1 – T5.

*3.5    Influence of the initial condition*

As shown in the example in Figure 7, the values for $E_\kappa$ evaluated from the TDS measurements using the fitting procedure were invariably much higher than $E_L$. A high $E_\kappa$ is necessary for the measured desorption flux at the beginning of the test to be zero. For low values of $E_\kappa$ the desorption flux at the beginning of the test is high. Thus a low value of $E_\kappa$ results in a high deviation between the theoretical curve and the experimental measurement at low temperatures. This is illustrated in Figure 9, which shows the desorption flux and sample temperature for T4 at a heating rate of 0.24 K·s$^{-1}$ plotted against time. The measured curve has been fitted assuming $E_\kappa = E_L$ and finding the optimum values for $N$, $E_\lambda$ and the initial H concentration. The calculated hydrogen flux is initially high and deviates markedly from the initial part of the measured curve.

The deviation between the theoretical and measured TDS curves at low temperatures implies that the shape of the curve, and the trapping parameters obtained from it, are dependent on the distribution of hydrogen in the sample at the onset of heating. A key source of variation in the TDS measurements is the effusion and redistribution of hydrogen between the end of the charging procedure and the onset of heating. This was investigated by calculating the TDS curves using the optimised values for $N$, $E_\lambda$ and the initial H concentration obtained by fitting the measured TDS curve in Figure 9 assuming $E_\kappa = E_L$ and delays of 150, 300 and 600 s before the onset of heating. The results are also shown in Figure 9. The horizontal axis is such that heating is first applied at 0 s. An increase in delay corresponded with a decrease in desorption flux at the onset of heating and a reduction in deviation from the measured curve. This suggests that the low initial desorption flux of the measured curves, and the high values of $E_\kappa$ obtained from them, is consistent with hydrogen effusion and redistribution between charging and the onset of heating.

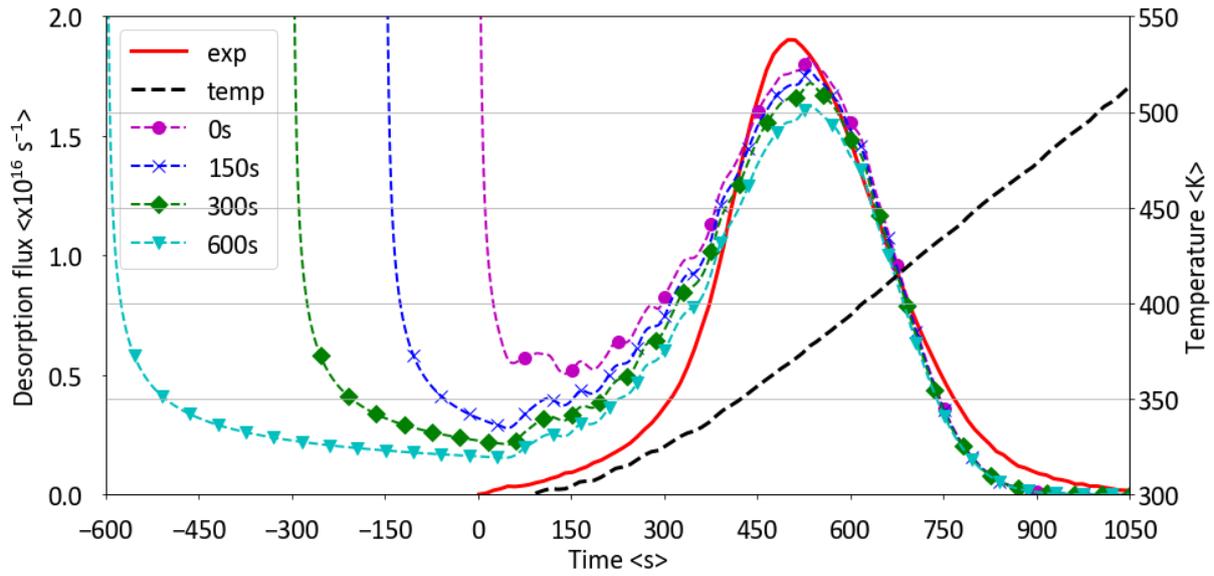

Figure 9 – TDS measurement for T4 at a constant heating rate of 0.24 K·s$^{-1}$ (solid red line) as well as theoretical curves obtained using assuming $N$ = 8.619×10$^{25}$ m$^{-3}$, $E_\kappa = E_L$ and $E_\lambda$ = 0.465 eV and delays of 0 s (●), 150 s (✕), 300 s (◆), and 600 s (▼) before the onset of heating. The horizontal axis is such that heating begins at 0 s. The measured temperature is shown with a dashed black line.

The measured TDS curve shown in Figure 9 was also fitted with the FE model by finding the optimum values of $E_\lambda$ and the initial H concentration assuming $E_\kappa = E_L$ and various values of $N$ between 5×10$^{24}$ m$^{-3}$ and 10$^{26}$ m$^{-3}$. Figure 10A shows the measured curve and two theoretical curves that were calculated assuming $N$ = 10$^{25}$ m$^{-3}$ and $N$ = 10$^{26}$ m$^{-3}$. For both theoretical curves there was no delay between the end of charging and the onset of heating. There is little deviation between the two theoretical curves except at temperatures below 350 K. In this temperature range both theoretical curves deviated significantly from the measured curve. The results of the fitting procedure for all values of $N$ between 5×10$^{24}$ m$^{-3}$ and 10$^{26}$ m$^{-3}$ are shown in Figure 10B. As in the previous case, there was a negative linear correlation between $\Delta E_T$ (or $E_\lambda$) and $\ln(N)$ with a gradient of -0.037 eV. This indicates that there is no unique combination of $N$ and $E_\lambda$ that will achieve a good fit with the measured curve assuming $E_\kappa = E_L$. The curve $\Delta E_T$ vs $\ln(N)$ is the same as that in Figure 7, albeit over a narrower range of $\ln(N)$. This suggests that the relationship between $\Delta E_T$ and $\ln(N)$ is independent of the initial conditions.

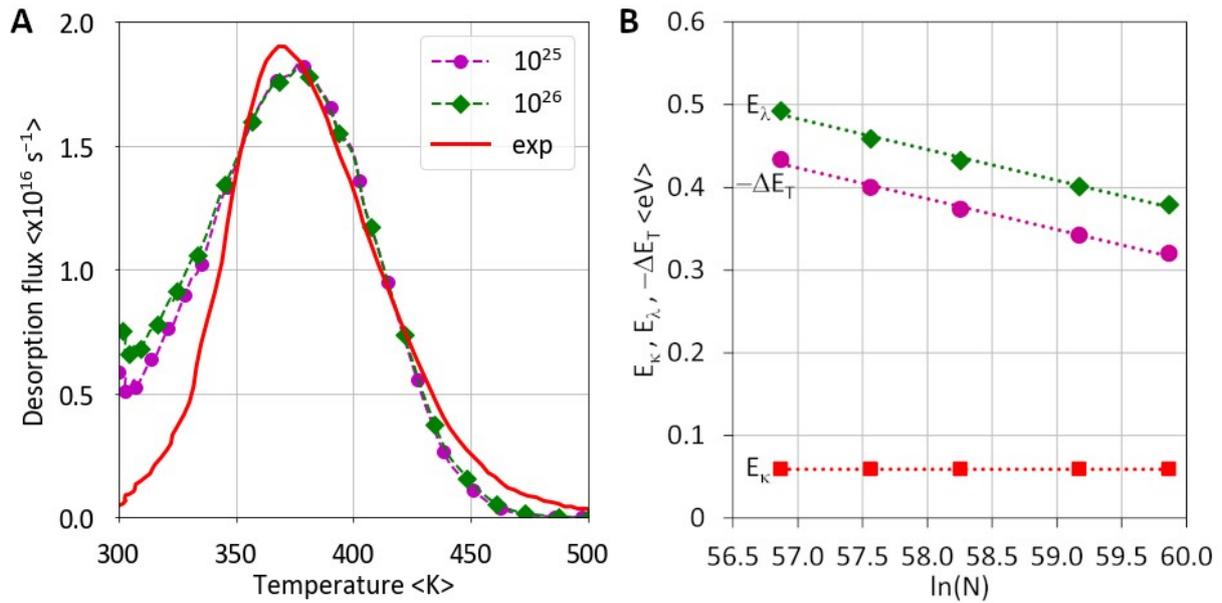

Figure 10 – (A) measured TDS curve for T4 at a constant heating rate of 0.24 K·s$^{-1}$ (solid red line) and theoretical curves obtained by assuming $E_\kappa = E_L$ and $N = 10^{25}$ m$^{-3}$ (●) or $N = 10^{26}$ m$^{-3}$ (◆) and finding the optimal value of $E_\lambda$ using the fitting procedure. (B) Optimised values $E_\lambda$ (◆), as well as $E_\kappa$ (■) and $\Delta E_T$ (●), obtained by fitting the numerical model to the measured curve shown in Figure 10A assuming $E_\kappa = E_L$ and various values of $N$ between $5 \times 10^{24}$ m$^{-3}$ and $10^{26}$ m$^{-3}$.

*3.6  Evaluation of trapping parameters using Kissinger equation*

The hydrogen trapping behaviour of the materials was also evaluated using the Kissinger equation (Equation 1). In this approach the activation energy for hydrogen desorption, $E_D$, is calculated based on the dependence of the temperature corresponding to the maximum desorption flux, $T_M$, on the heating rate. Figure 11 shows TDS measurements for T4 for various heating rates in the range 0.12 – 0.4 K·s$^{-1}$. With increasing heating rate there is an increase in $T_M$ and the maximum desorption flux.

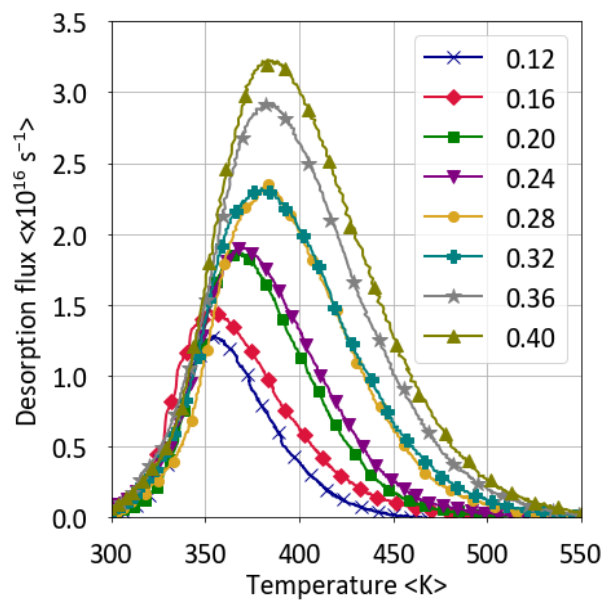

Figure 11 - TDS measurements for samples of material T4 measured at various heating rates (shown in K·s$^{-1}$).

Preliminary evaluation of $E_D$ using the Kissinger equation was performed by assuming a single type of trap. In this instance $T_M$ was taken as the temperature corresponding to the maximum desorption flux in the measured TDS curve. The plots of $\ln(\varphi/T_M^2)$ with respect to $1/T_M$ and the resulting values of $E_D$ are given in the supplementary materials. For T1, T4 and T5, $E_D$ was in the range 0.28 – 0.32 eV. For T2 and T3, $E_D$ was in the range 0.16 – 0.21 eV. This is inconsistent with the presence of fine (Ti,Cr)C particles in T2 and T3.

Subsequent analysis of $E_D$ was performed by assuming two types of traps. This involved deconvolving the experimental curves into two gaussian peaks, each with a height at least half that of the measured curve, as shown in Figure 12. The plots of $\ln(\varphi/T_M^2)$ with respect to $1/T_M$ for the two types of traps, denoted I and II, are shown in Figure 13. The corresponding values of $E_D$ are given in Table 2. For all materials $E_D$ for both types of traps was in the range 0.23 – 0.36 eV. There was no correlation between the $E_D$ and the presence or size of (Ti,Cr)C particles.

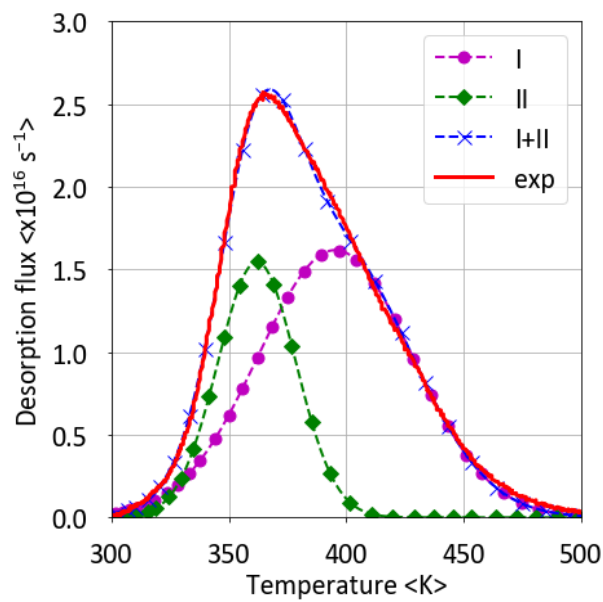

Figure 12 – TDS measurement of T2 heated at 0.28 K·s$^{-1}$. The experimental measurement ('exp') has been deconvolved into two gaussian curves (I and II) with peak heights at least half that of the measured curve.

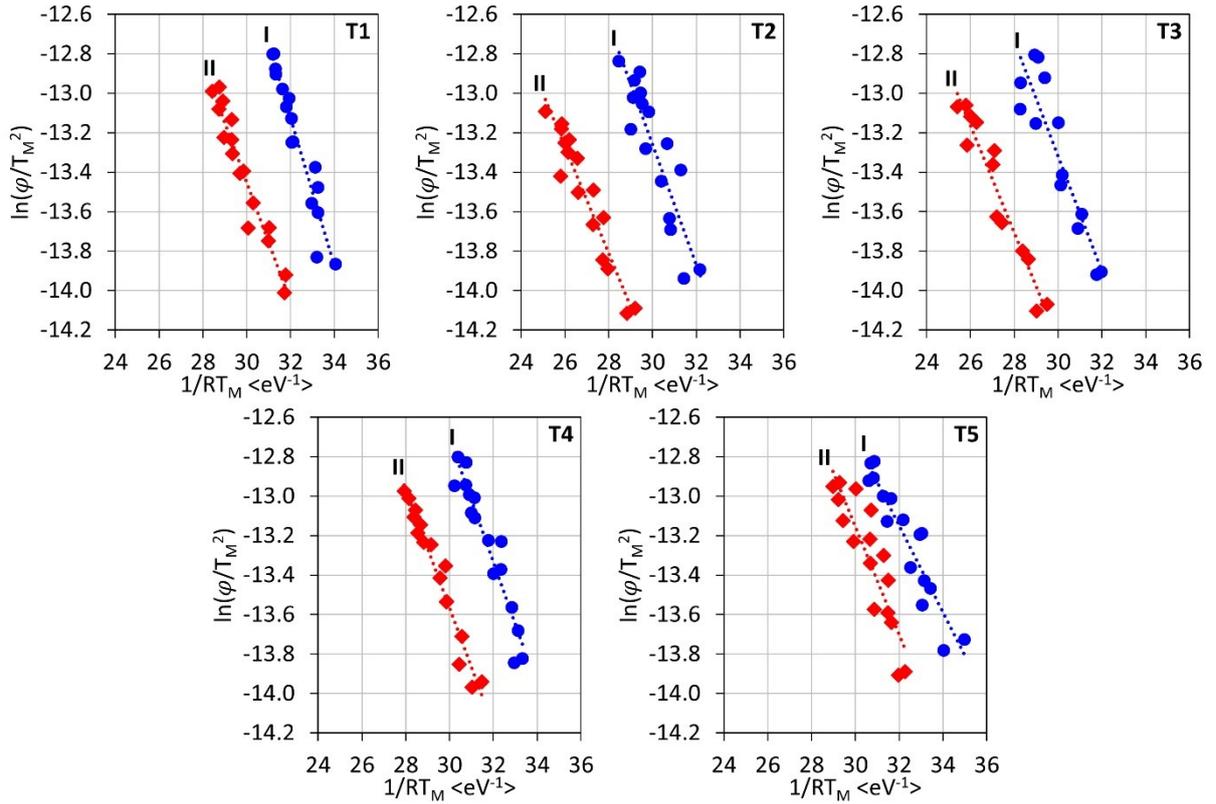

Figure 13 – $\ln(\varphi/T_M^2)$ plotted against $1/RT_M$ for all TDS measurements where the experimental curve was deconvolved into two peaks (I and II) as shown in Figure 12.

Table 2 – Activation energy for hydrogen desorption, $E_D$, derived from the results in Figure 13.

| Material | Trap | $E_D$ (eV) | $R^2$ |
|---|---|---|---|
| T1 | I | 0.37 | 0.92 |
| T1 | II | 0.30 | 0.94 |
| T2 | I | 0.31 | 0.80 |
| T2 | II | 0.27 | 0.92 |
| T3 | I | 0.29 | 0.83 |
| T3 | II | 0.27 | 0.93 |
| T4 | I | 0.31 | 0.91 |
| T4 | II | 0.30 | 0.95 |
| T5 | I | 0.22 | 0.88 |
| T5 | II | 0.27 | 0.79 |

## 4    Discussion

Qualitative evaluation of the EP and TDS measurements showed that (Ti,Cr)C particles strongly effect hydrogen diffusion in ferritic steels, with the influence of (Ti,Cr)C particles greatly dependent on their size. Particles less than 5 nm, such as those in T2 and T3, slowed hydrogen diffusion markedly. This was manifested by an increase in the time to reach steady state for the EP measurements and an increase in the maximum desorption flux and the corresponding temperature for the TDS measurements. Larger particles, such as those in materials T4 and T5, had little or no effect on hydrogen diffusivity. The dependence of trapping capacity on particle size is consistent with that reported by previous workers for TiC particles [64-67]. In the present study the hydrogen diffusivity of T1, which did not contain (Ti,Cr)C particles, was similar to that of T4 and less than that of T5. This is consistent with the role of dislocations as hydrogen trap sites; due to its bainitic microstructure the dislocation density of T1 would be higher than that of the other materials. For TDS measurements all hydrogen was desorbed at temperatures less than 600 K. There was no evidence of trapping sites with high binding energies that would be activated only at high temperatures, such as those associated with vacancies in the core of TiC particles [85]. This implies that traps associated with (Ti,Cr)C particles have a relatively low activation energy. However, it is not possible to attribute these traps to any particular site due to inconsistencies in the trap binding energies reported in the literature [85] and the difficulty of interpreting the results.

The measured TDS curves for all materials were characterised by a low desorption flux at the onset of heating. This is consistent with previous reports for BCC steels [38,54,55,75], but not consistent with theoretical TDS curves calculated using the McNabb-Foster model assuming a low energy barrier. Comparison of the theoretical TDS curves in Figure 7 with those in Figure 10 suggests that the maximum desorption flux and the corresponding temperature is dependent on $N$ and $\Delta E_T$, whereas the initial part of the curve is dependent on $E_\kappa$ and $E_\lambda$. If $E_\lambda$ is low, then desorption of hydrogen in traps may occur low temperatures. Thus good agreement with the initial part of the TDS curve can only be achieved by assuming a high $E_\lambda$. Since $\Delta E_T$ is independent of the initial part of the curve, a high $E_\lambda$ implies a high $E_\kappa$. In the examples shown in Figure 8, $E_\kappa$ was greater than 0.38 eV whilst $E_\lambda$ was greater than 0.87 eV. This was also true for T1, which did not contain (Ti,Cr)C particles. In contrast, it is often assumed in the literature [41,48,49,100,101] that $E_\kappa$ is equal to $E_L$ (0.059 eV [91]). This discrepancy may be partly due to the effusion and redistribution of hydrogen between the end of the charging process and the onset of heating; Figure 9 shows that an increase in delay results in a decrease in the initial desorption flux. Other factors, such as hydrogen trapping at the surface of the sample, may also affect the initial desorption flux but have not been investigated in this work. The values obtained for $E_\kappa$ and $E_\lambda$ should therefore be treated with caution. The relationship between $N$ and $\Delta E_T$ is not affected by the initial part of the curve or the values of $E_\kappa$ and $E_\lambda$ and may be considered a true reflection of the diffusion and trapping behaviour of the material.

A key outcome of this work is that the individual trapping parameters $N$, $E_\kappa$ and $E_\lambda$, as well as $\Delta E_T$, cannot be determined by fitting EP and TDS measurements using the McNabb-Foster model. Parameters obtained through the fitting procedure are not unique; the measurements may be fitted with multiple combinations of parameters. For both types of measurements there exists a negative linear correlation between $\Delta E_T$ and $\ln(N)$. The values for $\Delta E_T$ and $\ln(N)$ for the EP and TDS measurements were colinear. This suggests that they are both a reflection of the bulk diffusion and trapping behaviour of the material. However, a means of describing and comparing the trapping characteristics of different materials is required. Previous analysis of EP measurements of BCC steels by Winzer *et al.* [42] showed that for isothermal conditions $N$, $\kappa$ and $\lambda$ are coupled such that $N\kappa/\lambda = C$, where $C$ is a constant. Substituting Equation 4 and Equation 5 into $N\kappa/\lambda = C$ and rearranging gives:

$$-\Delta E_T = -RT\ln(N) + RT\ln(CN_L) \qquad \text{Equation 10}$$

In the present study EP measurement were performed at 303 K, whereas for TDS measurements the peak diffusion flux occurred at 350 – 400 K. In this temperature range -RT is equal to -0.026 – -0.034 eV. This is consistent with the gradient of the curve through $\Delta E_T$ vs $\ln(N)$ for the results in this study. Thus negative linear correlation between $\Delta E_T$ and $\ln(N)$ is consistent with the relationship $N\kappa/\lambda = C$ under isothermal conditions.

The results obtained using the Kissinger equation in this study were highly uncertain. This uncertainty was due to the dependence of the activation energy for hydrogen desorption, $E_D$, on the assumed number of traps. Evaluation of the TDS measurements assuming a single type of trap (i.e., by taking $T_M$ as the temperature corresponding to the maximum desorption flux of the measured TDS curves) indicated that $E_D$ for T2 and T3 was lower than that for the other materials. This is inconsistent with the presence of fine (Ti,Cr)C particles in T2 and T3. Evaluation of the measurements assuming two types of traps (i.e., by deconvolving the measured curves into two gaussian peaks) indicated that $E_D$ was the range 0.28 – 0.32 eV for all materials. It should be noted that the assumed number of traps is arbitrary; the measured TDS curves could be deconvolved into any number of gaussian curves. It is possible that different values for $E_D$ would be obtained by assuming a larger number of traps. The values obtained using the Kissinger equation were also inconsistent with those obtained using the fitting procedure. It should be noted that $E_D$ corresponds to $E_\lambda$. Assuming $E_\kappa = E_L$, an $E_D$ of 0.28 – 0.32 eV (see Table 2) corresponds to an $\Delta E_T$ of 0.22 – 0.26 eV. This is below the values of $\Delta E_T$ shown in Figure 8. For these reasons the values obtained using the Kissinger equation should be considered highly uncertain.

## 5     Summary

- The hydrogen trapping behaviour of ferritic steels containing (Ti,Cr)C particles was investigated using EP and TDS measurements. Steels with different sizes of (Ti,Cr)C particles were fabricated in the laboratory by varying the coiling temperature.
- The presence of (Ti,Cr)C particles less than 5 nm in size significantly slowed the rate of hydrogen diffusion in ferritic steels; larger particles had little or no effect. For TDS measurements all hydrogen was desorbed at temperatures less than 600 K. This is consistent with hydrogen trapping at dislocations and at the interface of (Ti,Cr)C particles, but not at vacancies in the core of the particles.
- It was not possible to evaluate the individual trapping parameters $N$, $E_\kappa$ and $E_\lambda$ (or $\Delta E_T$) by fitting EP and TDS measurements with a FE model based on the McNabb-Foster equations. The measurements could be fitted with multiple combinations of parameters. There was a negative linear correlation between $\Delta E_T$ and $\ln(N)$. This is consistent with previous work by Winzer et al. [42], which showed that $N\kappa/\lambda$ is constant for isothermal conditions.
- TDS measurements were evaluated using the Kissinger equation assuming one or two different types of traps. In the case of two traps, $\Delta E_T$ was in the range 0.22 – 0.26 eV for all materials. For some materials, the results were dependent on the number of traps and are, therefore, considered highly uncertain.

**Acknowledgements**

The author wishes to thank the technical staff at thyssenkrupp Steel Europe AG for carrying out the experiments in this report as well as Dr. Rémi Delaporte-Mathurin at the Massachusetts Institute of Technology for his assistance with FESTIM.

**Supplementary materials**

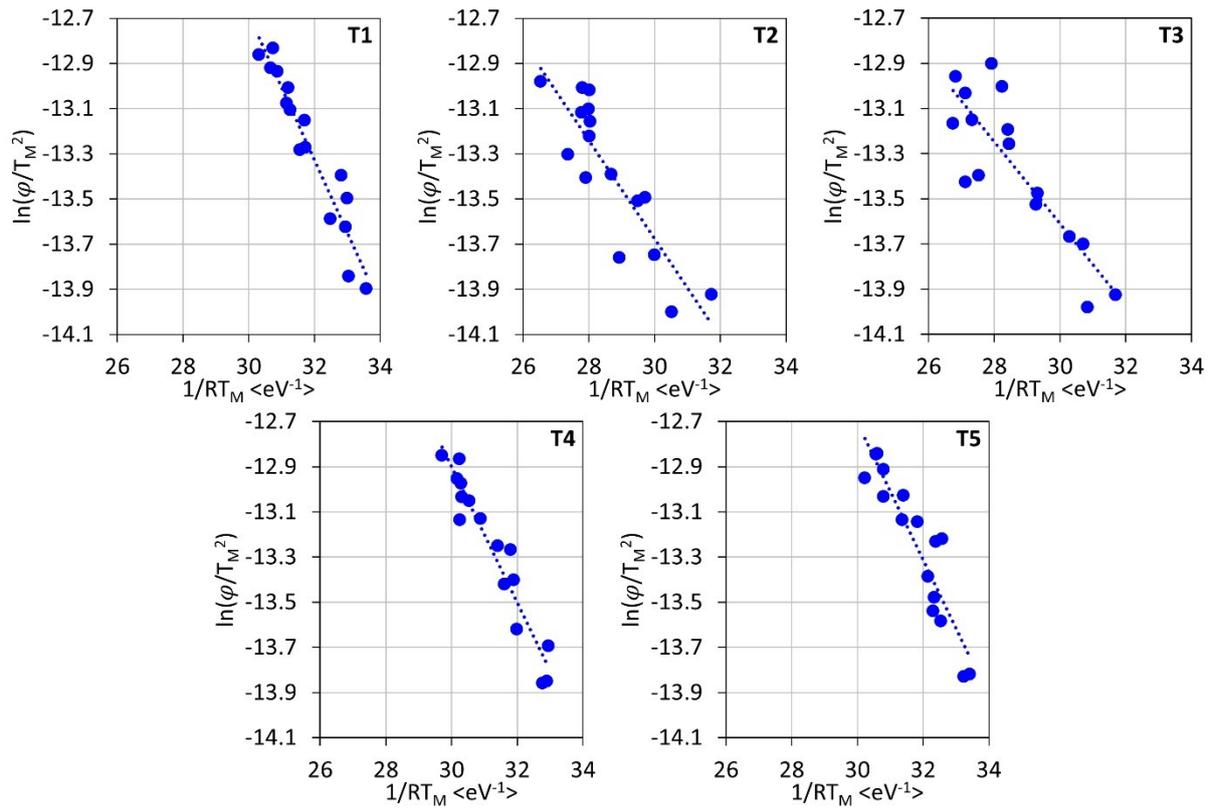

Figure 14 – $\ln(\varphi/T_M^2)$ plotted against $1/RT_M$ for all TDS measurements where $T_M$ was taken as the temperature corresponding to the maximum desorption flux in the experimental TDS curve.

Table 3 – Activation energy for hydrogen desorption, $E_D$, derived from the results in Figure 14.

| Material | $E_D$ (eV) | $R^2$ |
|---|---|---|
| T1 | 0.32 | 0.92 |
| T2 | 0.22 | 0.77 |
| T3 | 0.18 | 0.72 |
| T4 | 0.3 | 0.92 |
| T5 | 0.3 | 0.85 |